\documentclass[12pt,preprint]{aastex}
\usepackage{aastexug}
\newcommand{\pks}{PKS~0637--752}

\usepackage{graphicx}
\usepackage{fancyheadings}
\usepackage{color}
\begin{document}

\title{A \emph{CHANDRA} SEARCH FOR X-RAY JETS IN REDSHIFT 6 QUASARS}
\author{D.A. Schwartz\altaffilmark{1},
 }
\altaffiltext{1}{Harvard-Smithsonian Center for Astrophysics.}

\email{das@head-cfa.harvard.edu}
\slugcomment{DRAFT, \today, das}

\begin{abstract}

We have searched for X-ray jets in the recent \emph{Chandra}
observations of three Sloan Digital Sky Survey (SDSS) quasars at
redshift z $\approx$ 6. All 3 quasars were detected in X-rays in these
relatively short observations. SDSS1030+0524 is not consistent with a
point source, and may be a gravitationally lensed system. We find a
possible jet-like feature 23$\arcsec$ from the quasar SDSS 1306+0356. We can
explain the emission by inverse Compton (IC) scattering off the Cosmic
Microwave Background (CMB), assuming that the intrinsic properties of
the system are similar to X-ray jets at z$<$1.  Deeper observations to
investigate the interpretation as a jet will be practical.

\end{abstract}


\section{INTRODUCTION}

The sub-arcsecond angular resolution of the \emph{Chandra} X-ray Observatory
has proven to have unique capability to resolve jets and hot-spots in
quasars and radio sources. This was shown in the very first pointed
observation \citep{Schwartz00, Chartas00}, at the z=0.652 quasar \pks\, for which \citet{Schwartz00} showed
that the X-rays could not plausibly arise from the synchrotron
mechanism. To explain the X-ray emission, \citet{Tavecchio00}, and
\citet{Celotti01} suggested that bulk relativistic motion of the jet
would allow the X-rays to be produced by inverse Compton (IC) radiation 
from the cosmic microwave background (CMB), while allowing the
magnetic fields and relativistic electrons to be near equipartition in
the jet rest frame.  \citet{Schwartz01,Schwartz02} noted that if observed
X-rays were produced by IC/CMB, whether or not relativistically
beamed, then such a source would be seen at
any larger redshift with the same surface brightness, and would be
resolved and detected by \emph{Chandra}. 

\citet{Fan01} have discovered the three most distant quasars known, at
redshifts 6.28, 5.99, and 5.82. A Director's Discretionary Program
(cf. http://cxc.harvard.edu/DDT/DDTobs\_info.html)
proposed by Brandt was carried out on 29 January, 2002 to obtain
``snapshot'' data for immediate public use in planning further studies
of distant quasars. We have used these data to search for X-ray jets
associated with those quasars.

\section{OBSERVATIONS OF THREE SDSS QUASARS AT REDSHIFT 6}
\label{sec:appearance}

The observations were carried out in the standard 6-chip ACIS-S
imaging mode, with a 3.2s frame time, and events telemetered in the
3$\times$3 faint mode format. We have selected for analysis only the data from
S3, with ASCA grades 0,2,3,4,6, and energy between 0.5 and 7 keV. At
lower energy the quantum efficiency and energy response is
increasingly uncertain, and at higher energies the background is
increasing steeply.  

Table~\ref{tab:sdssObs} summarizes the observations. The observing
time is taken from the number of frames, times the 3.2s frame
time. For the core, we tabulate the measured counts in a 1$\farcs$23
radius about the quasar position (5 pixel diameter), giving
$\sim$~95\% encircled energy fraction at 1.5 keV \citep{Jerius00}. We
convert from a measured counting rate in the 0.5--7 keV band to an
unabsorbed, measured flux in that band assuming a power law spectrum
of energy index 0.7, and the galactic hydrogen column density. This is
approximately 6.0$\times$10$^{-12}$ ergs cm$^{-2}$ s$^{-1}$ per count
s$^{-1}$. The factor changes about $\pm$ 1.5\% for the range of column
densities of the three objects, 2.1 to 4.4 $\times$10$^{20}$cm$^{-2}$,
and changes by about $\pm$ 20\% for indices from 0.4 to 1.2. Poisson
statistics on the few detected counts dominate the uncertainty. The
final two columns convert to rest frame luminosity in the 2--10 keV
band\footnote{We use $H_0 = \rm 65\, km\, s^{-1}\, Mpc^{-1}$ and a
flat universe with $\Omega_0 = 0.3$, $\Omega_{\Lambda} = 0.7$, and
q$_0$ = -0.55 throughout.}, including the K-correction for the assumed
spectrum.

\begin{deluxetable}{ccccccccc}    
\footnotesize
\tablewidth{0pt}
\tablecaption{Observations of the SDSS quasars at redshift 6\label{tab:sdssObs} }
\tablecolumns{9}
\tablehead{
\colhead{} & \colhead{} & \colhead{} & \colhead{Core} & \colhead{Core}
& \colhead{Jet} & \colhead{Jet} & \colhead{} & \colhead{} \\*
\colhead{Name\tablenotemark{a}} & \colhead{redshift\tablenotemark{a}} &
\colhead{Time(Ks)} & \colhead{Counts} & \colhead{Flux\tablenotemark{b}}
& \colhead{Counts\tablenotemark{c}} & \colhead{Flux\tablenotemark{b,c}} &
\colhead{L$_{\mathrm{core}}$\tablenotemark{d}} &
\colhead{L$_{\mathrm{jet}}$\tablenotemark{c,d}} 
}

\startdata

SDSSp J083643.85+005453.3  &5.82 &5.686 &21 &2.2
& $<$ 6.3 & $<$0.66 &2.3 &$<$0.70 \\
SDSSp J103027.10+052455.0 &6.28  &7.942&6 & 0.45 &$<$6.3 &$<$0.48  &0.55 &$<$0.59\\
SDSSp J130608.26+035626.3 &5.99  &8.160& 16 & 1.2 & 7
 &0.51	 & 1.3 &0.57  \\

\enddata
\tablenotetext{a}{\citet{Fan01}}
\tablenotetext{b}{Received 0.5--7 keV flux, unabsorbed,  in units of
10$^{-14}$ergs cm$^{-2}$ s$^{-1}$ }
\tablenotetext{c}{Detection is significant, but identification as a
jet is not certain}
\tablenotetext{d}{Rest frame 2 --10 keV luminosity in units of 10$^{45}$ergs s$^{-1}$}

\end{deluxetable}

\begin{figure}[h]
	\plotone{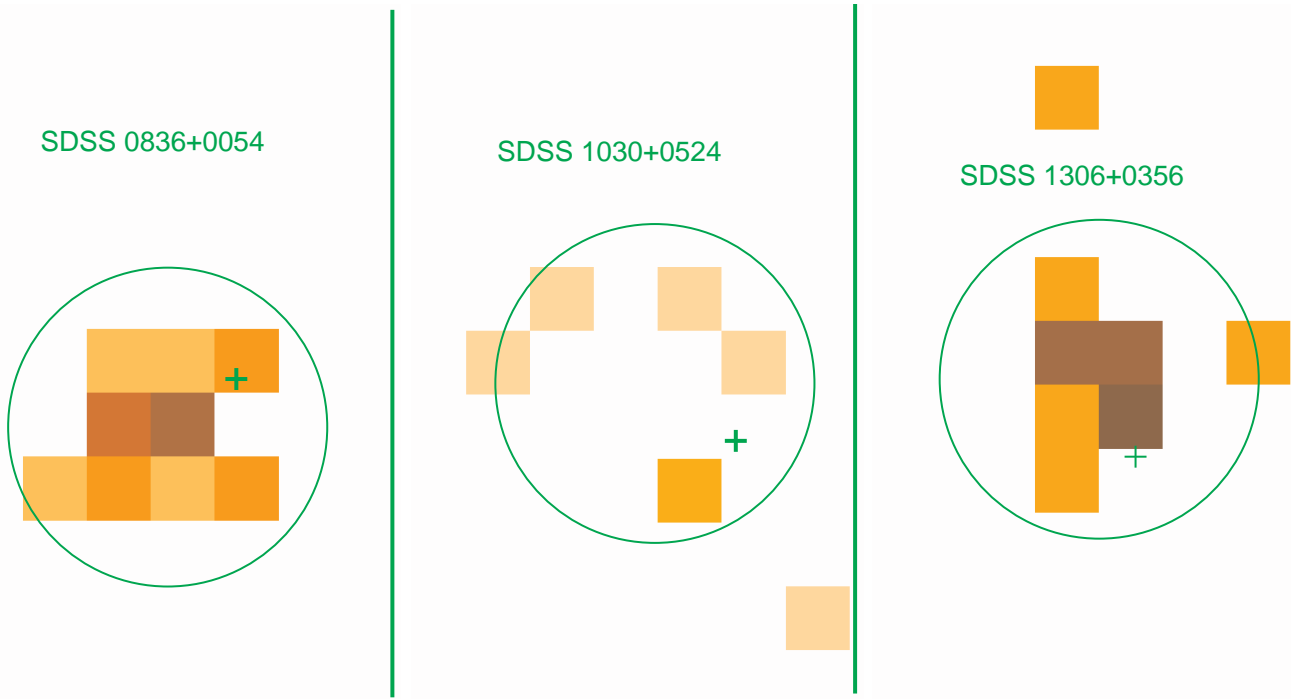} 
 \caption{\label{fig:corefield}
Detection of X-ray emission from the three SDSS quasars at
 z$\approx$6. The plus signs show the positions given by
 \citet{Fan01}. The X-ray centroids are all consistent within the
 absolute astrometric accuracy of \emph{Chandra}. In the 1$\farcs$23 radius extraction circles we
 find 21, 6, and 14 counts respectively, for SDSS 0836, SDSS 1030, and SDSS
1306. The pixels displayed have from 1 to 6 counts. The fields shown are 3$\farcs$5 $\times$ 5$\farcs$4. The X-rays
 are binned into 0$\farcs$4913 ACIS pixels.} 
\end{figure}

Figure~\ref{fig:corefield} shows the detection of each quasar.  The
plus sign gives the true position of the quasar \citep{Fan01}. The
X-ray centroids are offset by 0$\farcs$65, 0$\farcs$76, and
0$\farcs$66, respectively for SDSS 0836, SDSS 1030, and SDSS
1306. These are within the astrometric performance of \emph{Chandra}
(cf. http://cxc.harvard.edu/cal/ASPECT/celmon/). Each quasar is
clearly detected.  The background rates are determined from large
rectangular regions to be 2.04, 2.56, and 2.92 $\times$10$^{-3}$
counts per pixel, respectively, so there is a negligible background
correction of 0.04 to 0.05 counts, which we do not make.  In the 19
pixel source extraction circles, the probability is less than
2.6$\times$10$^{-5}$ of getting 3 or more counts, so all the
detections are highly significant.  (For a serendipitous source
anywhere in the field, one only needs 4 counts to have less than a 1\%
probability of a chance occurrence.) SDSS 0836 and SDSS 1306 are
clearly consistent with being point sources.

The 6 photons from SDSS 1030+0524 are highly unlikely to be from a
single point source.  They are all further than 0\farcs68 from their
centroid. That distance is the 79\% encircled energy radius, so the
probability is 8.6$\times\,10^{-5}$ that all six photons would fall
outside this radius. There is a 4.7\% chance that one is due to
background, but less than 0.13\% chance that two or more are
background counts. If we discard the most distant photon and recompute
the X-ray centroid, one falls at
the 55\% encircled radius and the other four are still outside the
80\% radius, and the probability of at least four so distant from the
centroid  is only 6.7$\times10^{-3}$.
Therefore either the source is extended, over about a 2\arcsec\   
diameter, or it is really two or more point sources. As discussed by
\citet{Wyithe02a,Wyithe02b}, there is a probability between 7\% and
30\% that any quasar at z=6 is gravitationally lensed, and this could be
confirmed with a followup observation of $>$50 ks.

\begin{figure}[h]
	\plotone{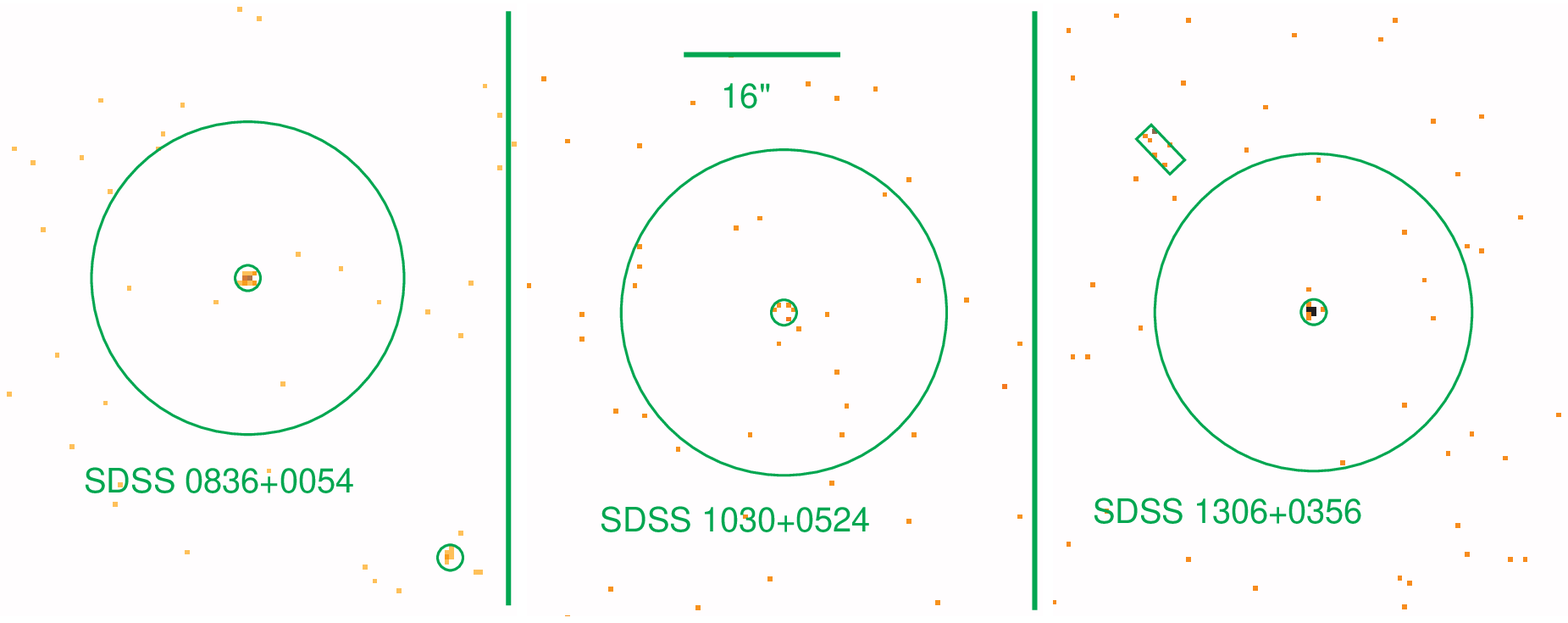} 
 \caption{\label{fig:jetfield}
Fields of 52$\arcsec\times$62$\arcsec$ around the SDSS quasars. The
 larger circles show a radius of 100 kpc, projected on the plane of the
 sky, at the distance of each quasar. These are 16$\arcsec$ to
 16$\farcs$6. The box to the NE of SDSS 1306
 has the 2$\arcsec$ full width of the telescope response, and points
 toward the quasar core. It contains 7 counts where 0.13 are expected
 due to background. We interpret this as a possible jet, but cannot
 exclude other combinations of sources and background. The 7 photons
 to the SW of SDSS 0836 clearly show the condensation expected for a
 point source.	} 
\end{figure}

Figure~\ref{fig:jetfield} shows the region surrounding each quasar.
For the cosmological parameters used here, 1$\arcsec$ corresponds to
6.26, 6.00 and 6.16 kpc for SDSS 0836, SDSS 1030, and SDSS 1306,
respectively.  The larger circles indicate a 100 kpc projected
distance from each quasar. There are no significant concentrations of
photons anywhere within these circles. In the annulus we expect 6.8,
9.3, and 10.0 background counts and observe 6, 14, and 8,
respectively. 

SDSS 1306+0356 has a significant X-ray feature 23$\farcs$3 to the NE
of the quasar, at a distance 143 kpc projected in the plane of the
sky. There are 7 counts in a box 5$\arcsec \times$2$\arcsec$ which
points toward the quasar. The significance of these as an X-ray source
is beyond doubt. We can arbitrarily place this box to include at least
one count.  Since there are only 0.13 background counts expected, the
chance of getting 6 additional counts in a 10 arcsec$^2$ box is only
5.8$\times$10$^{-9}$, or only 1.7$\times$10$^{-4}$ of occurring in any
10 arcsec$^2$ box within 5$\arcmin$ of the quasar. There is no visible
object at this position in the digital sky survey. The FIRST radio
survey gives an upper limit of 0.93 mJy at 1.4 GHz at this position
\citep{White97}. Interpretation as an IC/CMB X-ray jet is entirely
reasonable, as we discuss below; however, there is no specific
evidence requiring such interpretation. Arguments as given above for
the core of SDSS 1030 rule out a single point source, with or without
attributing any of the 7 counts to background. The spatial
distribution of counts cannot rule out two point sources and a
background count.  \citet{Giacconi01} give about 200 sources
deg$^{-2}$ above the flux of 5$\times$10$^{-15}$ observed from the
jet, so there is a 5\% chance of one such source falling within
1$\arcmin$ of the quasar.  The chance occurrence of an extended source
at this flux level is at least 10 times smaller, $<$0.5\% (see
\citet{Bauer02}). Because a longer X-ray observation is needed to
obtain definitive information, and because we cannot rule out the
possibility of foreground sources correlated with each other, we do
not attempt a more exhaustive discussion of the probabilities.  As a
newly discovered X-ray source, we designate this as
CXOU~130609.31+035643.5.

In Table~\ref{tab:sdssObs} we set a limit to the flux of putative jets
in SDSS 0836 or SDSS 1030 by noting that we do not find 3 or more
photons in any similar 10 arcsec$^2$ box pointing toward the
core. If there were a jet from which we expected 6.3 or more counts,
then there would be less  than a 5\% probability to fail to detect 3
or more, and we adopt this as our limit.

\begin{figure}[h]
	\plotone{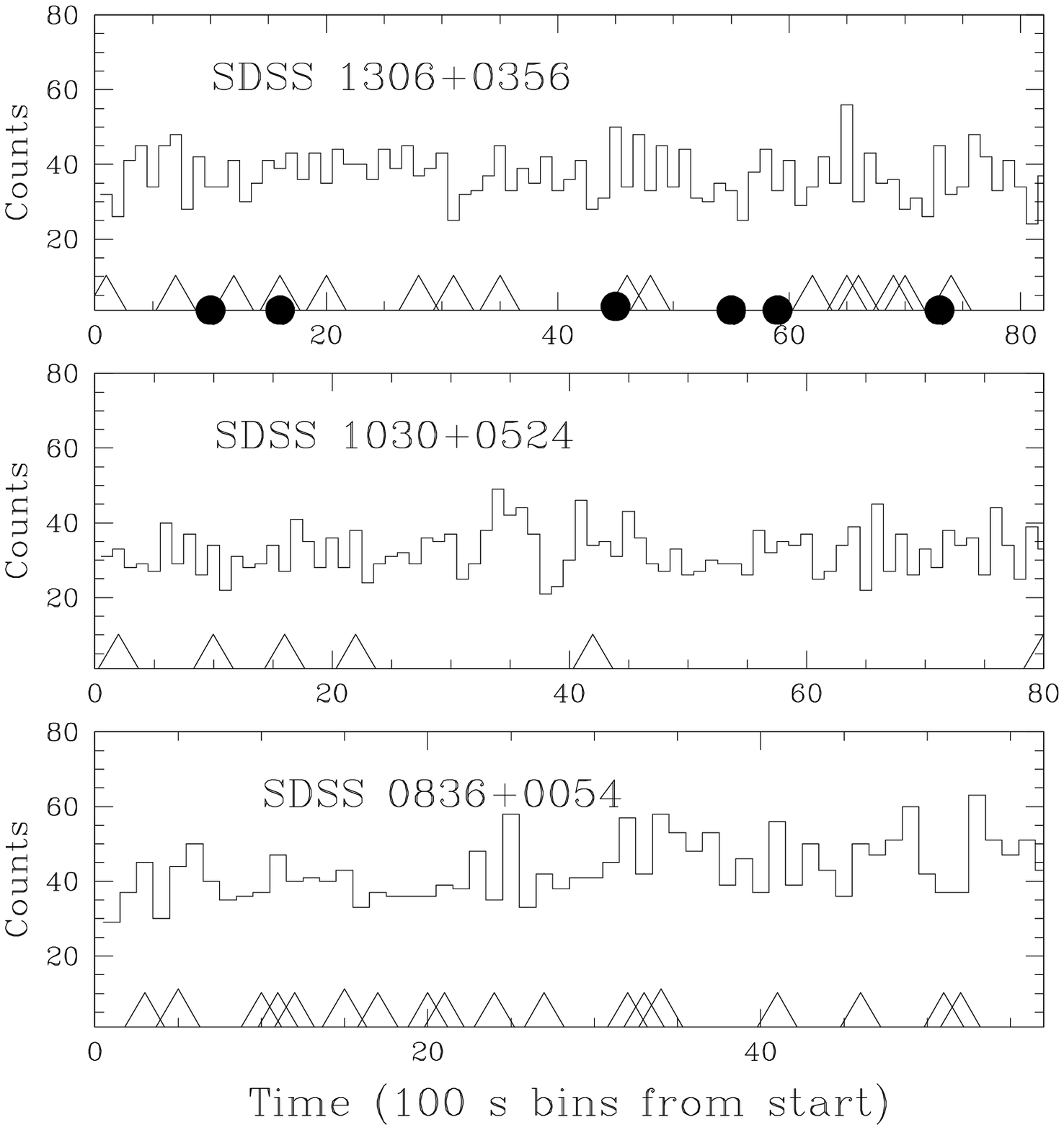} 
 \caption{\label{fig:times}
The solid histogram gives the arrival times of all 0.5 to 7 keV counts in the
 field, binned in 100s intervals from the start of each
 observation. (The final partial bin is not shown.) The upward
 triangles show the arrival time of each photon from the quasar
 extraction circle. The solid dots for SDSS 1306 show the arrival
 times of photons in the box which is a candidate as an X-ray jet.} 
\end{figure}

Because of the small numbers of photons, we need to give careful
consideration to systematic effects. 
Figure~\ref{fig:times} shows that the background was steady during
these observations, and the arrival times of the quasar and jet counts
are consistent with being uniform throughout the observations. No two
counts were registered from the same physical pixel. 

\begin{figure}[h]
         \plotone{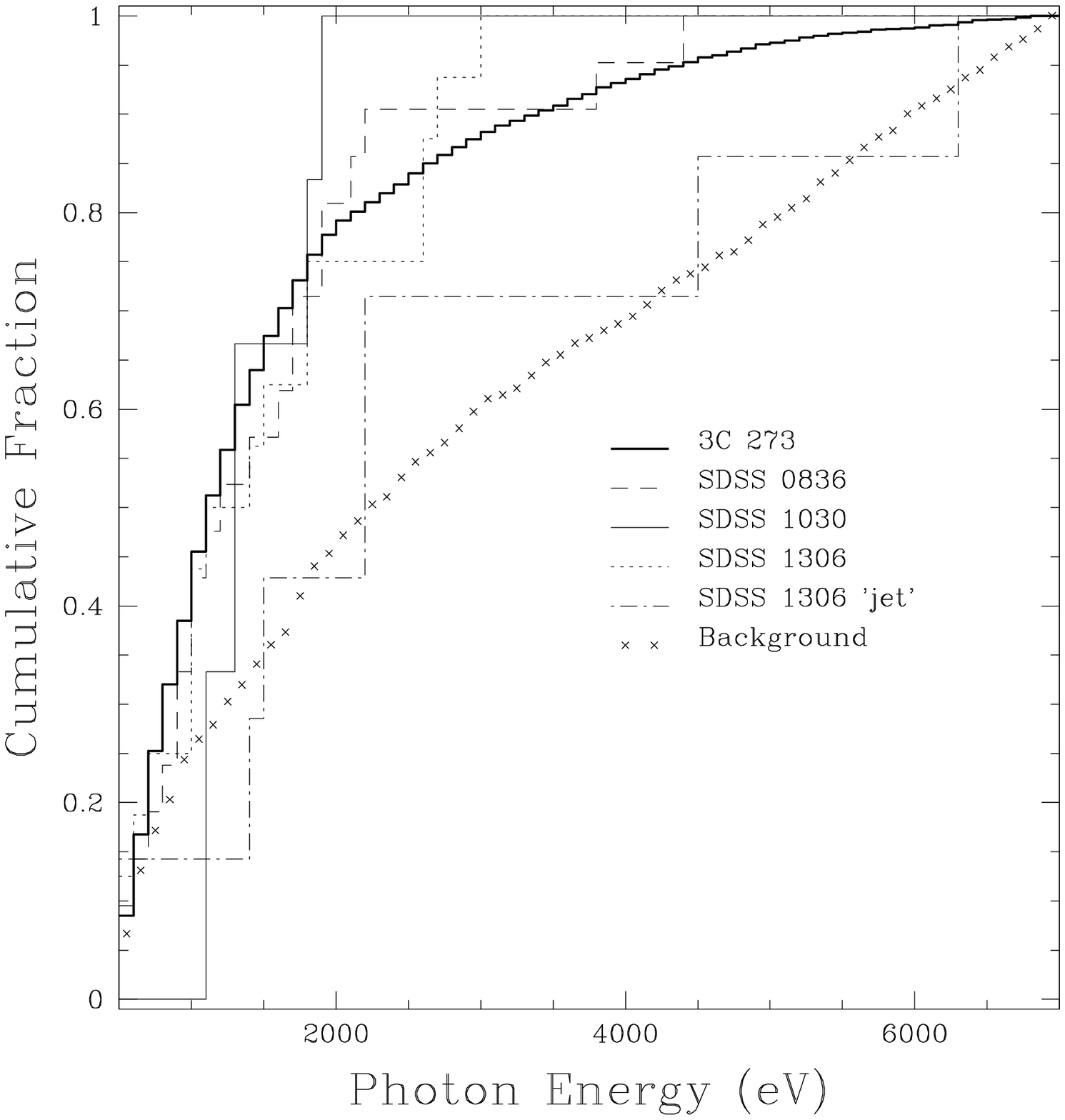}
\caption{\label{fig:energytest}Integral probability distributions of
the photon energies from the SDSS redshift 6 quasars.  All photons are
binned by 100 eV, from 500 to 7000 eV. A KS test allows
all spectra to be consistent with the $\alpha$=0.65 index of 3C273,
shown as the heavy curve and based on 3063 photons. The jet spectrum
is quite  consistent with that of the background counts, shown as the
small crosses, while the quasar spectra have only a small probability
of consistency with the background spectrum.}
\end{figure}

To investigate any peculiarities in the distribution of photon
energies, we perform a KS test of the energy distributions of the
photons from each quasar and from the jet feature against the core
spectrum of 3C 273. We might expect them \emph{a priori} to have a
non-thermal spectrum of energy index roughly $\alpha\sim$0.7.
Figure~\ref{fig:energytest} shows the cumulative distributions.  The
solid heavy line is the core spectrum of 3C 273, determined from 3063
photons in the readout streak in OBSID 1712 \citep{Marshall01}. If
SDSS 0836, SDSS 1030, SDSS 1306, and the jet all had the same spectrum
as 3C~273, we would expect deviations as large as actually observed
98\%, 16\%, 96\%, and 10\% of the time, respectively. Thus we cannot
reject this hypothesis for any of them. The crosses plot the
distribution of 763 background photons from a source free region of
the SDSS 1306 observation. Such large deviations as observed would be
expected only 0.3\%, 5.6\%, 2.8\%, and 90\% of the time, for the same
respective sources. Thus the quasar core spectra are very distinct
from that of the background, while the putative jet spectrum is
consistent with the harder background spectral shape.

We compute $\alpha_{\mathrm{ox}}$ according to the original definition
\citep{Tananbaum79},
\begin{equation}
\label{eq:alfox}
 \alpha_{\mathrm{ox}} =-\log(f_{x}/f_{o})/\log(\nu_{x}/\nu_{o}), 
\end{equation}
where $f_{x}$ is the X-ray
flux density at 2 keV, $\nu_{x}$= 4.8$\times$10$^{17}$ Hz, and $f_{o}$
is the optical flux density at 2500 \AA,
$\nu_{o}$=1.2$\times$10$^{15}$ Hz. The flux densities are computed in
the rest frame. We take the AB$_{1450}$ magnitudes from \citet{Fan01} and extrapolate from
1450\AA\,  to 2500\AA\, assuming a power law flux
$\nu^{-0.5}$ \citep{Richstone80,VandenBerk01}. We find
$\alpha_{\mathrm{ox}}$ of 1.66, 1.79, and 1.65, respectively, for SDSS
0836, SDSS 1030, and SDSS 1306. These values are consistent with the
range found by \citet{Vignali01} for \emph{Chandra} quasars at z$>$4,
but about 0.1 larger numerically than the mean found by \citet{Kaspi00} for
\emph{ROSAT} z$>$4 quasars. We note that an extraction circle larger
than 25$\arcsec$ applied to SDSS 1306+0356 would lead to a decrease in
$\alpha_{\mathrm{ox}}$ by 0.06, due to inclusion of flux from 
CXOU~130609.31+035643.5.

\section{CAN CXOU 130609.31 BE AN X-RAY JET FROM SDSS 130608.26?}

The feature discovered by \emph{Chandra} to the NE of SDSS 1306 can be
explained as a jet emitting by inverse Compton scattering on the
CMB. No specific evidence prohibits a synchrotron mechanism, but such
emission would imply much more severe energetic and lifetime
constraints, while the IC/CMB process implies that the source will
appear to have the same surface brightness at any redshift
\citep{Schwartz02}. The ratio of synchrotron (radio) emission to IC
(X-ray) emission from a population of electrons
\citep{Jones65,Felten66} is
\begin{equation}
\label{eq:densityratio}
 \frac{L_{R}}{L_{X}}=\frac{\mathrm{H}^2/(8 \pi)}{\rho_{CMB}}.
\end{equation}
Since the energy density of microwave photons at the quasar is
$\rho_{CMB}$ = aT$^4$ = 7.56$\times$10$^{-15}$
T$_{0}^4\,(~1~+~z~)^4$\\=10$^{-9}$ergs cm$^{-3}$ for the microwave
temperature of T$_0$ = 2.728 K \citep{Fixsen96} and redshift z=5.99,
$\rho_{CMB}$ will dominate any magnetic field of strength less than
160$\mu$G. But from the radio flux upper limit, we can derive a rest
frame 100 MHz to 100 GHz radio luminosity of less than
3$\times$10$^{44}$ ergs s$^{-1}$, compared to the rest frame 3.5--49
keV X-ray luminosity of 1.3$\times$10$^{45}$ ergs s$^{-1}$, and deduce
that H$\lesssim$80$\mu$G from equation~\ref{eq:densityratio}.

  Electrons with
$\gamma \approx$ 1000 will broadly scatter the CMB photons to observed
energies around 1 keV. Their lifetime will be 2.1$\times$10$^{12}$/($\gamma$
(1+z)$^4$) years. Taking the volume corresponding to a cylinder
5$\arcsec$ long and 1$\arcsec$ diameter at the source distance,
1.1$\times$10$^{68}$cm$^{3}$, gives a required energy density of
3.2$\times$10$^{-10}$ ergs cm$^{-3}$ in relativistic electrons
emitting the X-rays. A 90 $\mu$G field would be in equipartition
with such an electron density. The magnetic field could
be a factor of a few smaller, due to uncertainties in the quantities
or the modeling assumptions, and since there is not yet a measured
radio flux.

 We would typically expect fields of only
tens of $\mu$G based on observations of X-ray jets to date
\citep{Harris02,Sambruna02}. Those observations have often required
that the jet be moving with a bulk Lorentz factor $\Gamma$ of at least a few,
(see \citet{Harris02,Sambruna02}), in order that the CMB energy
density be enhanced in the rest frame of the jet by a factor
$\Gamma^2$, and still preserve equipartition of the magnetic field and
relativistic particle energy densities. Because of the large redshift
we need not invoke relativistic beaming to explain the present
observations. However, it is certainly allowed, as long as the jet is
pointed within an angle $\arccos(\sqrt(\frac{\Gamma -1}{\Gamma+1}))$
of our line of sight, so that the apparent flux is not diminished. 

That the jet flux is about 40\% of the quasar core flux in this case
is consistent with the ratios of roughly 1\% commonly observed in
X-ray jets\footnote{But note that we so far do not have extensive,
systematic X-ray surveys of jet properties.} if we assume that the jet
magnetic field is approximately 30 $\mu$G. In that case, if this
object were at a redshift z=2, the core flux would be enhanced a
factor of 14. The jet surface brightness would remain the same due to
compensating factors of (1+z)$^4$ in the microwave energy density and
the cosmological dimming, and the solid angle would be about a factor
of 2 smaller, for a net decrease in the jet to core ratio to about 1.5\%. At redshifts less than 2, if it were not relativistically beamed,
the CMB energy density would be less than the magnetic field energy
density and the intrinsic X-ray emissivity of the jet would no longer
be affected by the CMB.

Although the putative jet is about 30 kpc long, projected on the sky,
there is an apparent gap of 130 kpc from the core to the jet, where
the surface brightness is a factor of at least two smaller. This could
be explained if the energy flux in this portion of the jet is
primarily in the form of protons or Poynting flux,
(cf. \citet{Harris02}), or if it is mildly relativistic but away from
our line of sight. In the latter case, the portion we see could be
isotropic radiation after the jet becomes sub-relativistic. 

\section{CONCLUSIONS}
All three z$\approx$6 quasars were detected in relatively short
\emph{Chandra} observations. Their X-ray to optical luminosity ratios
are median values for high redshift, high luminosity quasars,
indicating that quasars at higher L$_{\mathrm{x}}$/L$_{\mathrm{O}}$,
which are known to exist, will be detectable at if they exist at larger
redshifts. Followup observations of all will be important to obtain
rough spectral information, to investigate if SDSS 1030+0524 is
gravitationally lensed, and to look more deeply for X-ray jets. In
particular, the candidate jet from SDSS 1306 could be confirmed, or
contradicted, by the spatial structure of 100 photons which would
obtain in a 100 Ks observation. Radio emission should be detectable at
$\gtrsim$100$\mu$Jy at 5 GHz, by scaling from \pks. This could be
detected in a few hour VLA BnA observation. If confirmed, this
will clearly indicate the possibility of detecting \emph{only} the
jet, and not core, X-ray emission from similar quasars at even larger
redshifts.

\acknowledgments This work was supported by NASA contract NAS8-39073
to the \emph{Chandra} X-ray Center.  I thank W. Forman, C. Jones,
A. Siemiginowska, H. Tananbaum, and B. Wilkes for discussion and
comments on the manuscript. A. Loeb pointed out the possibility that
SDSS 1030 could be a gravitationally lensed system.  This research
used the NASA Astrophysics Data System Bibliographic Services, and the
NASA/IPAC Extragalactic Database (NED) which is operated by the Jet
Propulsion Laboratory, California Institute of Technology, under
contract with the National Aeronautics and Space Administration.

\end{document}